\documentclass[]{scrartcl}% insert '[draft]' option to show overfull boxes
\usepackage{amsmath}
\usepackage{multirow}
\usepackage{multicol}
\usepackage{caption}
\usepackage{subcaption}
\usepackage{wrapfig}
\usepackage{amsmath}
\usepackage{amsfonts}
\usepackage{graphicx}
\usepackage[none]{hyphenat}
\usepackage{float}
\usepackage{calc}
\usepackage{algorithm}
\usepackage{algorithmic}

 \title{Pseudospectral Model Predictive Control under Partially Learned Dynamics}

 \author{
  Manan Gandhi%
  	\thanks{Ph.D Student, School of Aerospace Engineering, Student Member}
  \ , Yunpeng Pan%
  	\thanks{Ph.D Candidate, Institute for Robotics and Intelligent Machines, Non-Member}
  \ and Evangelos Theodorou%0
  	\thanks{Assistant Professor, School of Aerospace Engineering, Non-Member}\\
  {\normalsize\itshape
   Georgia Institute of Technology, Atlanta, Georgia}\\
 }

 \newcommand{\explain}[2]{\underbrace{#1}_\text{#2}}

\begin{document}

\maketitle

\begin{abstract}
Trajectory optimization of a controlled dynamical system is an essential part of autonomy, however many trajectory optimization techniques are limited by the fidelity of the underlying parametric model. In the field of robotics, a lack of model knowledge can be overcome with machine learning techniques, utilizing measurements to build a dynamical model from the data. This paper aims to take the middle ground between these two approaches by introducing a semi-parametric representation of the underlying system dynamics. Our goal is to leverage the considerable information contained in a traditional physics based model and combine it with a data-driven, non-parametric regression technique known as a Gaussian Process. Integrating this semi-parametric model with model predictive pseudospectral control, we demonstrate this technique on both a cart pole and quadrotor simulation with unmodeled damping and parametric error. In order to manage parametric uncertainty, we introduce an algorithm that utilizes Sparse Spectrum Gaussian Processes (SSGP) for online learning after each rollout. We implement this online learning technique on a cart pole and quadrator, then demonstrate the use of online learning and obstacle avoidance for the dubin vehicle dynamics.
\end{abstract}

\section*{Nomenclature}
\begin{multicols}{2}
\begin{tabbing}
  XXXX \= \kill% this line sets tab stop
  $\mathbf{x}$ \> $ \mathcal{R}^{n \times 1} $ state vector  \\
  $\mathbf{u}$ \> $ \mathcal{R}^{m \times 1} $ control vector \\
  $t$ \> scalar time variable \\
  $J$ \> scalar cost  \\
  $Q$ \> $ \mathcal{R}^{n \times n} $ constant state weighting  \\
  $R$ \> $ \mathcal{R}^{m \times m} $ constant control weighting  \\ 
  $\psi$ \> cost to go to the next discretized state \\
  $\mathcal{L},g$ \> cost function utilized for running cost \\
  $\Psi (t_f)$ \> terminal cost \\ 
  $h_s $ \> state constraints \\ 
  $h_c $ \> control constraints \\ 
  $\mathbf{C}$ \> boundary constraints \\ 
  $\mathbf{Z}$ \> Training data for Gaussian Process \\ 
  $\mathbf{Y}$ \> Training observations for GP \\
  $\mathbf{z}^*$ \> test input for GP \\
  $\mathbf{y}^*$ \> prediction from GP \\
  $\mathbf{K}$ \> kernel function for GP \\
  $\mathbf{A}$ \> Approximate kernel for Sparse Spectrum GP \\
  $\mathbf{w}$ \> Random Fourier feature weights for SSGP \\
  $O_r$ \> Obstacle radius \\
  $B_r$ \> Body radius \\
  $U_r$ \> Uncertainty radius
 \end{tabbing}
\end{multicols}

\section{Introduction}
Guidance and navigation in aerospace has the capability to leverage high fidelity physics based models when generating a trajectory or designing a controller. These models can be expensive to evaluate or sometimes unavailable; thus an increasing trend in robotics and machine learning is to use a data-driven approach. In these frameworks, the measurements taken during system identification and control tasks are used to build a dynamical model for the system. The drawbacks to purely data-driven approaches are that they often require multiple, repeated trials on the real system which is not always practical. Thus, we hope to leverage the advantages of both techniques through the use of a semi-parametric representation of the dynamics: a combination of a physics based model and a Gaussian Process.

Gaussian processes are a data efficient, non-parametric regression technique that can be utilized to learn any nonlinear function, including dynamics \cite{wang2005gaussian}. The method is non-parametric because there is no explicit, parameter-based characterization of the function, just a kernel function between training data and new inputs. In addition to data-efficiency, a Gaussian Process is a probabilistic inference method, meaning that we can determine both the mean and variance of our prediction based on a test input. This has lead to a variety of applications for a Gaussian process. One example is learning the model behind human motion capture data, where each pose is up to 50 dimensions \cite{wang2008gaussian}.  An extension to this method is to use a semi-parametric approach to learn the inverse dynamics for a 7-DoF BARRET WAM arm \cite{nguyen2010using}. Here they characterize the error dynamics between their model and rollouts to the system with a Gaussian Process. In the context of control, Gaussian processes have been used to characterize the uncertainty of a dynamical model for a learning based robust control algorithm \cite{berkenkamp2015safe}. Finally, in the context of trajectory optimization, there has been work to combine a Gaussian process an algorithm known as Differential Dynamic Programming. As a result, the new algorithm, coined Probabilistic Differential Dynamic Programming, is capable of completely learning the dynamics of a system, while simultaneously performing trajectory optimization on it\cite{pan2014probabilistic}.

Gauss pseudospectral optimal control is our trajectory optimization algorithm of choice because of three reasons: natural incorporation of state and control constraints, integration accuracy of the Gauss quadrature, and efficiency of dynamics function calls. We utilize GPOPS-II for an implementation of the Gauss Pseudospectral optimal control method \cite{GPOPS}, as well as a in-house python implementation that naturally allows for data driven dynamics. We use a model predictive approach to incorporate feedback into our algorithm and improve robustness against disturbances \cite{gandhi2016comparison}. MPC also sets up a framework to incorporate online learning. The rest of the paper is organized as follows. First we introduce the mathematical background with the problem formulation, and presentation of Gaussian processes and Gauss Pseudospectral Optimal Control. Next we summarize the base algorithm and present the simulations on the Cart Pole and Quadrotor systems. Finally we introduce an online learning algorithm and compare perform the trajectory optimization tasks for even more uncertain systems, as well as for an obstacle avoidance problem.

\section{Mathematical Background}

First we present the problem and necessary background material on Gaussian Processes and Gauss Pseudospectral optimal control. The Gaussian process derivation can be found in detail by the Gaussian Process text by Rasmussen \cite{rasmussen2006gaussian}, and Gauss Pseudospectral has its roots in Hunting's thesis \cite{Huntington} and has been further extended by Patterson and Rao with GPOPS-II \cite{GPOPS}.

\subsection{Problem Formulation}

We would like to minimize a user-defined cost function $J$ subject to dynamics $f_t(\mathbf{x},\mathbf{u},t)$ and inequality constraints $h_s(\mathbf{x})$ and $h_c(\mathbf{u})$.
\begin{eqnarray}
\centering \label{ProblemCost}
 &\underset{\mathbf{u}}{\text{min}} ~J(\mathbf{x}(t),\mathbf{u}(t)) =  \underset{\mathbf{u}}{\text{min}} ~\bigg( \explain{\psi(\mathbf{x}(t_f),t_f)}{Cost to go} + \explain{ \int_{t_0}^{t_f}\mathcal{L}(\mathbf{x}(t),\mathbf{u}(t),t)dt}{Running Cost} \bigg) \\ \nonumber
&  \text{subject to} ~~\dot{\mathbf{x}}(t) = f_t(\mathbf{x},\mathbf{u},t) \\
&  \qquad \qquad  ~~~0 \geq h_s(\mathbf{x}) \\
&  \qquad \qquad  ~~~0 \geq h_c(\mathbf{u}) 
\label{ProblemDyn}
\end{eqnarray}

\noindent $   \Psi(\mathbf{x}(t_f),t_f) $ is the terminal cost  and $ \mathcal{L}(\mathbf{x}(t),\mathbf{u}(t),t) $ is the running cost. 

$f_t$ represents the true dynamics, a function that we do not have complete knowledge of. Instead, we assume that $f_t$ can be represented as the sum of two functions that we do know: a parametric physics model $f_p$ and a non-parametric Gaussian Process $f_{err}$. In our current formulation we will only utilize the predictive mean of the Gaussian process as part of our dynamics, there are other frameworks that incorporate a predictive variance \cite{pan2014probabilistic}$^,$\cite{miller2009coordinated} .

\begin{eqnarray}
f_t(\mathbf{x}(t),\mathbf{u}(t)) &=& \explain{f_p(\mathbf{x}(t),\mathbf{u}(t))}{Parametric physics model} + \explain{f_{err}(\mathbf{x}(t),\mathbf{u}(t))}{Non-parametric Gaussian Process}
\end{eqnarray}

Next we go into further detail about Gaussian Processes and how they fit into our framework. 

\subsection{Gaussian Processes}

A Gaussian Process (GP) is defined as ``a collection of random variables, any finite number of which have a Gaussian distribution'' \cite{rasmussen2006gaussian} . We can collect $N$ pairs of state-control pairs, $\mathbf{Z} =  \{ (\mathbf{x}_1,\mathbf{u}_1 ) , ... ,(\mathbf{x}_N,\mathbf{u}_N ) \}$, and the corresponding outputs from our true dynamics $\mathbf{Y} = \{ (\mathbf{y}_1) , ... ,(\mathbf{y}_N) \}$ ,when $\mathbf{y}_i = f_t(\mathbf{x}_i(t),\mathbf{u}_i(t)) - f_p(\mathbf{x}_i(t),\mathbf{u}_i(t)) ~~ i = 1,..,N$. Classical Gaussian processes are scalar-valued function approximators, meaning that we will have an independent Gaussian process for each dimension of  $\mathbf{y}_i$. Methods exist for correlated vector outputs of Gaussian Processes \cite{alvarez2011kernels} , however for our purposes we have not found significant evidence for including correlated vector outputs. The output of a Gaussian process is completely defined as a normal distribution with some predictive mean and variance. The joint distribution of the observed output (from our test cases) and the output corresponding to a test case $\mathbf{z^*} = (\mathbf{x^*},\mathbf{u^*}) $ can be written as the following equation:

\begin{eqnarray}
p ~\bigg( \begin{matrix}\mathbf{~Y~}\\ \mathbf{~y}^* \end{matrix} \bigg) \sim 
\mathcal{N} \bigg( ~ 0 , \bigg[ \begin{matrix} \mathbf{K}(\mathbf{Z},\mathbf{Z}) + \sigma_n^2 \mathbf{I} & \mathbf{K}(\mathbf{Z},\mathbf{z}^*) \\ \mathbf{K}(\mathbf{z}^*,\mathbf{Z}) & \mathbf{K}(\mathbf{z}^*,\mathbf{z}^*) \end{matrix} \bigg] ~\bigg)
\end{eqnarray} 

Note that the joint distribution is characterized with a user defined Kernel function $\mathbf{K}(\mathbf{z}_i,\mathbf{z}_j)$. We consider the kernel function corresponding to the squared exponential or Gaussian kernel: 

\begin{eqnarray}
\mathbf{K}(\mathbf{z}_i,\mathbf{z}_j) = \sigma_s^2 \exp( -\frac{1}{2} (\mathbf{z}_i - \mathbf{z}_j )^T \mathbf{W} (\mathbf{z}_i - \mathbf{z}_j ) ) + \sigma_n^2
\end{eqnarray}

where $\sigma_s, \sigma_n$, and $\mathbf{W}$ are hyper-parameters that are must be trained with the data. $\sigma_s$ is a measure of the variance of the data itself, $\sigma_n$ is a measure of noise of the observations, and $\mathbf{W}$ is a diagonal matrix containing the characteristic length scales of each input. We see that the covariance function between outputs is written in terms of the inputs, implying that we are making some assumptions about the smoothness of the underlying function we are trying to learn. The text by Rasmussen \cite{rasmussen2006gaussian} shows that this particular covariance function corresponds to a Bayesian linear regression model with an infinite number of basis functions.  

Next we can condition the joint prior Gaussian distribution on the observations to obtain the following:

\begin{eqnarray}
\mathbf{y}^* | \mathbf{z}^*,\mathbf{Z},\mathbf{Y} &\sim& \mathcal{N} \big(\mathbf{K}(\mathbf{z}^*, \mathbf{Z}) (\mathbf{K}(\mathbf{Z},\mathbf{Z})+ \sigma_n^2 \mathbf{I})^{-1} \mathbf{Y},
\mathbf{K}(\mathbf{z}^*, \mathbf{z}^*) - \mathbf{K}(\mathbf{z}^*, \mathbf{Z}) (\mathbf{K}(\mathbf{Z}, \mathbf{Z})+ \sigma_n^2 \mathbf{I})^{-1} \mathbf{K}(\mathbf{Z}, \mathbf{z}^*) \big)
\end{eqnarray}

This is the one step prediction of our output $\mathbf{y}^*$ given our test input $\mathbf{z}^*$ and our training data $(\mathbf{Z},\mathbf{Y})$. In some applications this prediction can be extended to a multi-step prediction, where we propagate the probabilistic dynamics, however, in our framework, we do not need to do this since the Gauss pseudospectral method approximates the dynamics at each collocation point separately, meaning we only perform a one step prediction. Note that performing a prediction requires inversion of the kernel function $\mathbf{K}(\mathbf{Z}, \mathbf{Z})+ \sigma_n^2 \mathbf{I}$. This inversion has computational complexity of $\mathcal{O}(N^3)$ where $N$ is the number of data points. Clearly using a Gaussian process in an online learning fashion is not scalable with respect to the number of data points, as you gather more data you must perform a larger and more difficult inversion at each update. There are techniques such as the inclusion of forgetting factors and local Gaussian Processes to improve the computational performance \cite{nguyen2009model} . Additionally we can perform online learning via an approximate Gaussian Process \cite{lazaro2010sparse}$^,$\cite{gijsberts2013real} . This is known as a sparse approximation to a Gaussian kernel or a Sparse Spectrum Gaussian Process.

\subsubsection*{Sparse Spectrum Gaussian Processes}

The key aspect of the Sparse Spectrum Gaussian Process (SSGP) is employing a random Fourier feature approximation of the kernel function \cite{rahimi2007random}. Based on Bochner's theorem, shift-invariant kernel functions can be represented as the Fourier transform of a unique measure \cite{rudin2011fourier}.

\begin{eqnarray}
\mathbf{K}(\mathbf{z}_i - \mathbf{z}_j) & = & \int_{\mathcal{R}^{N}} e^{i\mathbf{\omega}^T(\mathbf{z}_i - \mathbf{z}_j)}p(\mathbf{\omega})d\mathbf{\omega} \\
									   & = & \mathbb{E}_\mathbf{\omega} [\mathbf{\phi}(\mathbf{z}_i)\mathbf{\phi}(\mathbf{z}_j)]
\end{eqnarray}

\noindent where $\mathbf{\phi}(\mathbf{z}) = [\cos(\mathbf{\omega}^T \mathbf{z}) \sin(\mathbf{\omega}^T \mathbf{z}) ] $. The term $ p(\mathbf{\omega}) $ is a probability distribution. The number of samples $r$ taken from the distribution $ p(\mathbf{\omega}) $ allows us to construct an unbaised approximate squared exponential function.

\begin{eqnarray}
\mathbf{K}(\mathbf{z}_i - \mathbf{z}_j) &\approx & \sum_{i = 1}^{r} \mathbf{\phi}_{\mathbf{\omega}_i}(\mathbf{z}_i)^T\mathbf{\phi}_{\mathbf{\omega}_i}(\mathbf{z}_j) \\
									  & = & \mathbf{\phi}(\mathbf{z}_i)^T\mathbf{\phi}(\mathbf{z}_j)
\end{eqnarray}

\noindent where

\begin{eqnarray}
\mathbf{\phi}_{\mathbf{\omega}}(\mathbf{z}) = \frac{\sigma_s}{\sqrt{r}} [\cos(\mathbf{\omega}^T \mathbf{z}) \sin(\mathbf{\omega}^T \mathbf{z}) ],~~~ \mathbf{\omega} \sim \mathcal{N}(0,\mathbf{W}^{-1})
\end{eqnarray}

Now that we have this efficient feature mapping, we can represent our underlying dynamics function $f_{err}(\mathbf{z})$ as a weighted sum of the feature functions. We assume a prior distribution of $\mathbf{w} \sim \mathcal{N}(0, \Sigma_p )$ on the weight matrix. The complete posterior distribution can be calculated as follows:

\begin{eqnarray}
\mathbf{y}^* | \mathbf{z}^*,\mathbf{Z},\mathbf{Y} &\sim & \mathcal{N}(\mathbf{w}^T\mathbf{\phi}(\mathbf{z}^*) ~,~ \sigma_n^2(1 + \mathbf{\phi}^T\mathbf{A}^{-1}\mathbf{\phi}))
\end{eqnarray}

\noindent where 

\begin{eqnarray}
\mathbf{\phi}(\mathbf{z}) &=& \frac{\sigma_s}{\sqrt{r}} \bigg[ \begin{matrix}  \cos(\mathbf{\Omega}^T\mathbf{z}) \\ \sin(\mathbf{\Omega}^T\mathbf{z})  \end{matrix} \bigg] \\
\mathbf{\Phi} &=& [ \phi(\mathbf{z}_1) , ... , \phi(\mathbf{z}_N) ] \\
\mathbf{A} &=& \mathbf{\Phi \Phi^T} + \sigma_n^2 \Sigma_p^{-1} \\
\mathbf{w} &=& \mathbf{A}^{-1} \mathbf{\Phi} \mathbf{Y} 
\end{eqnarray}

With both standard Gaussian Processes and Sparse Spectrum Gaussian Processes, the derivative with respect to the input can be computed analytically, thus enabling a significant cost savings when utilizing trajectory optimization techniques. The whole purpose behind introducing SSGP's was to incorporate online learning, which we will introduce in the next section. 

\subsubsection*{Online model learning}

Incremental online learning provides a way to update the SSGP. The computational complexity of has become $ \mathcal{O}(Nr^2+r^3) $ where $r$ is the number of random features of the SSGP. To update the weights $\mathbf{w}$ as new samples come in, we store the upper triangular Cholesky factor $\mathbf{A} = \mathbf{R^TR}$ \cite{gijsberts2013real}. With each new sample, $\mathbf{R}$ is updated with computational complexity $ \mathcal{O}(r^2) $. We also utilize a forgetting factor $\lambda \in (0,1)$ to assign a lower weight to data that is no longer relevant (particularly in the case of time varying systems).

\begin{eqnarray}
\mathbf{A}_{i+1} &=& \lambda \mathbf{A}_i + (1 - \lambda) \mathbf{\phi}(\mathbf{z_{i+1}})\mathbf{\phi}(\mathbf{z_{i+1}})^T \label{Aupdate} \\ 
\mathbf{Y}_{i+1} &=& \lambda \mathbf{y}_i + (1 - \lambda) \mathbf{\phi}(\mathbf{z_{i+1}})\mathbf{y}_{i+1} \label{yupdate}
\end{eqnarray}

The updates to $\mathbf{R}$ are rank-1, and the weights $\mathbf{w}$ can be computed by

\begin{eqnarray}
\mathbf{w} = (\mathbf{R^TR}^{-1})\mathbf{y}
\end{eqnarray}

\subsection{Gauss Pseudospectral Optimal Control}
We present an overview of pseudospectral methods based off our previous work \cite{gandhi2016comparison}. Pseudospectral optimal control is in a class of algorithms known as "direct" methods. The general outline \cite{Huntington} is to do the following: first, we will convert the dynamic system into a problem with a finite set of variables and transform the differential constraints into algebraic constraints with spectral approximation for the derivatives. This will allow us to solve with a non-linear program (NLP) solver. Second, we will solve this finite dimensional problem with the NLP and optimize the parameters with which we approximated the original system. Third, we can assess the accuracy of our approximation utilizing the costate and the Karush-Kuhn-Tucker (KKT) conditions. The advantage of spectral differentiation for smooth problems is that it exhibits a fast convergence rate. The Gauss Pseudospectral Method is described in much further detail in the referenced thesis by Huntington \cite{Huntington}.

Let us begin with a few equations to describe the problem. The Pseudospectral methods are defined with the Transformed Continuous Bolza problem. The problem formulation given by Equations \eqref{ProblemCost} and \eqref{ProblemDyn} must be put into the following form:

\begin{eqnarray}
J = \Psi(\mathbf{x}(\tau_0),t_0,\mathbf{x}(\tau_f),t_f) + \frac{t_f-t_0}{2} \int_{\tau_0}^{\tau_f}g(\mathbf{x}(\tau),\mathbf{u}(\tau),\tau;t_0,t_f)d\tau \label{Cost}
\end{eqnarray}

This cost will be minimized subject to the following constraints.
\begin{equation}
\frac{d\mathbf{x}}{dt} = \frac{t_f-t_0}{2}\mathbf{f}_t(\mathbf{x}(\tau),\mathbf{u}(\tau),\tau;t_0,t_f) \label{dyn}
\end{equation}
\begin{equation}
\mathbf{C}(\mathbf{x}(\tau),\mathbf{u}(\tau),\tau;t_0,t_f) \leq \mathbf{0} \label{bound}
\end{equation}
\begin{eqnarray}
\tau = \frac{2t}{t_f-t_0}-\frac{t_f+t_0}{t_f-t_0} \label{time}
\end{eqnarray}

The function $g$ represents the integrated cost similar to the purpose of $\mathcal{L}$ in DDP. Equation (\ref{dyn}) is the dynamic constraint to the problem. Equation (\ref{bound}) is the boundary condition at the final state, similar to our constraints $h_c$ and $h_p$ in the original problem. Equation (\ref{time}) shows the time transformation required for this derivation of pseudospectral optimal control. Typically, we require a fixed time interval such as $[-1,1]$, thus such a transformation can be used that is still valid with free initial and final times $t_0$ and $t_f$.

\subsubsection*{Collocation Point Search}
Collocation points are where we set the approximation to the nonlinear dynamic constraints equal to the true function across the interval. Below is the equation for the Legendre-Gauss points. The points are defined as the roots of Equation (\ref{LG}). Equation (\ref{LG}) is the $K^{th}$ degree Legendre polynomial.

\begin{equation}
P_K(\tau) = \frac{1}{2^KK!}\frac{d^K}{d\tau^K}\big[(\tau^2-1)^K \big] \label{LG}
\end{equation}

This choice of points mitigates the Runge phenomenon and improves accuracy towards the boundaries.

\subsubsection*{State, Control, and Costate Approximation}

Next we will approximate the state, control, and the costate. Note that we do not have to use the same collocation points and basis functions for each state, control, and costate, but we do use the same process.

\begin{equation}
\mathbf{x}(\tau) \approx \mathbf{X}(\tau) = \sum_{i=1}^K \mathcal{M}_i(\tau)\mathbf{X}(\tau_i) \label{state}
\end{equation}

\begin{equation}
\mathcal{M}_i(\tau) = \prod_{j=1,j\neq i}^{K} \frac{\tau-\tau_j}{\tau_i-\tau_j} = \frac{g(\tau)}{(\tau-\tau_i) \dot{g}(\tau)} \label{Lag}
\end{equation}

\begin{equation}
g(\tau) = (1+\tau)P_k(\tau)
\end{equation}

Equation (\ref{state}) represents the state approximation in terms of a set of Lagrange interpolating polynomials. Equation (\ref{Lag}) represents the actual polynomials, where the $P_K$ is the Legendre Polynomial. 

\subsubsection*{Integral Approximation via Gauss Quadrature}

Within the dynamic constraints and cost function we also have integrals that must be approximated. One way to do this is via a Gauss quadrature. The generic form for this is in Equation (\ref{GQ}).

\begin{equation}
\int_{a}^{b}f(\tau)d\tau \approx \sum_{i=1}^{K}w_if(\tau_i) \label{GQ}
\end{equation}

The points $\tau_k$ are the quadrature points on the interval $[-1,1]$ and the weights $w_i$ are the quadrature weights. Equation (\ref{qweight}) is for finding these weights.

\begin{equation}
w_i = \int_{-1}^{1}\mathcal{M}_i(\tau)d\tau = \frac{2}{(1-\tau_i^2)[\dot{P}_k(\tau_i)]^2},  (i = 1,...,K) \label{qweight}
\end{equation}

\subsubsection*{Algrebriac Representation of Dynamics}

The next step is to convert the differential equation constraint on the optimization problem into an algebraic constraint. This is done in Equation(\ref{DtoA})

\begin{equation}
\dot{\mathbf{x}}(\tau_k) \approx \dot{\mathbf{X}}(\tau_k) = \sum_{i=1}^{K}\dot{\mathcal{M}}(\tau_k)\mathbf{X}(\tau_i) = \sum_{i=1}^{K}D_{ki}\mathbf{X}(\tau_k), (k = 1,...,K) \label{DtoA}
\end{equation}

The differentiation matrix which is given in Equation (\ref{Diff}).
\begin{equation}
D_{ki} = \begin{cases}
\frac{(1+\tau_k)\dot{P}_K(\tau_k)+P_K(\tau_k)}{(\tau_k-\tau_i)\big[(1+\tau_i)\dot{P}_K(\tau_i)+P_K(\tau_i) \big]} &\text{if } k\neq i\\
\\
\frac{(1+\tau_i)\ddot{P}_K(\tau_i)+2\dot{P}_K(\tau_i)}{2\big[ (1+\tau_i)\dot{P}_K(\tau_i)+P_K(\tau_i) \big]} & \text{if } k = i

\end{cases} \label{Diff}
\end{equation}

We can now implement this algebraic representation in terms of a residual function by equating the derivative of our approximation to the vector field.

\begin{equation}
\mathbf{R}_k = \sum_{i=1}^{K}D_{ki}\mathbf{X}(\tau_k) - \frac{t_f-t_0}{2}\mathbf{f}(\mathbf{X}(\tau_k),\mathbf{\mathbf{U}}(\tau_k),\tau_k;t_0,t_f) = \mathbf{0},  (k = 1,...,K) \label{res}
\end{equation}

\subsubsection*{Gauss Pseudospectral Discretization of the Continuous Bolza Problem}

After these transformations, we collect terms for our new problem. We want to minimize the following cost function Equation (\ref{DBP}) subject to the algebraic collocation constraints in Equation (\ref{AC}). We also have the quadrature constraint in Equation (\ref{QC}), and path constraint in Equation (\ref{PC}).

\begin{equation}
J = \Psi(\mathbf{X}_0,t0,\mathbf{X}_f,t_f) + \frac{t_f-t_0}{2}\sum_{k=1}^{K}w_kg(\mathbf{X}_k,\mathbf{U}_k,\tau_k;t_0,t_f) \label{DBP}
\end{equation}

\begin{equation}
\mathbf{R}_k = \sum_{i=0}^{K}D_{ki}\mathbf{X}_i-\frac{t_f-t_0}{2}\mathbf{f}(\mathbf{X}_k,\mathbf{U}_k,\tau_k;t_0,t_f) = \mathbf{0}  (k = 1,...,N) \label{AC}
\end{equation}

\begin{equation}
\mathbf{R}_f = \mathbf{X}_f-\mathbf{X}_0-\frac{t_f-t_0}{2}\sum_{k=1}^{K}w_k\mathbf{f}(\mathbf{X}_k,\mathbf{U}_k,\tau_k;t_0,t_f) = \mathbf{0} \label{QC}
\end{equation}

\begin{equation}
\mathbf{C}(\mathbf{X}_k,\mathbf{U}_k,\tau_k;t_0,t_f) \leq \mathbf{0} \label{PC}
\end{equation}

The procedure to solve is as follows.

\begin{enumerate}
\item Choose the initial and final times, then choose your basis functions and use the Legendre polynomials to get the collocation points.
\item Translate the dynamic system into the the discretized and approximated version of the Continuous Bolza Problem. Also translate the constraints into the equivalent forms.
\item Solve the finite-dimensional problem with a Non-linear Program Solver.
\item Check the accuracy of the problem by the costate approximation, and then run again with increased collocation points, ideally moving towards the optimum.
\end{enumerate}

One program utilized to implement this method is called GPOPS II \cite{GPOPS}. Note we directly substitute $f_p + f_{err} = f_t$ into the optimizer. We also implement this technique in python, the main difference in our programming being that we allow the representation of our error dynamics $f_err$ to be an arbitrary function with arbitrary analytical derivatives that are based on data. This way machine learning techniques can be more efficiently implemented with this powerful framework. Also of note is that we can naively incorporate the variance of our Gaussian Process prediction in the instance of obstacle avoidance. Since Pseudospectral Optimal Control allows for continuous nonlinear constraints that are satisfied explicitly in the problem formulation, we can write these as obstacles and perform path planning and control simultaneously with a probabilistic bound.

\section{Results}

For the following simulation results, we implement GP's with GPOPS-II on two primary systems: a cart pole and quadrotor model. On the cart pole we introduce unmodeled viscous friction and learn the error dynamics through an untuned feedback controller. For the quadrotor model, we have a $ 10\% $ mass difference between the true model and the parametric model, and we hope to absorb the resulting error into the Gaussian process.

\begin{table}[H]
\centering
\label{Problem}
\caption{Gaussian Process Example Problem Settings}
\begin{tabular}{l | l | c | c| c}
Problem & Parameters & True & Nominal & Cost Function\\ \hline \hline
\multirow{6}{5em}{Cart Pole}& Cart Mass $m_c$ (kg)		&  0.5 & 0.5  &
						 \\ & Link Mass  $m_p$ (kg)		& 0.5  & 0.5  & $\int_{t_0}^{tf}(R\mathbf{u}^2)dt $
					 	 \\ & Link Length  $l$ (m)		& 0.6  & 0.6  & 
					 	 \\ & Ground damping $b_1$ (N/s) & 0.01  &  0   &
					 	 \\ & Pivot damping $b_2$ (Nm/s) & 0.01  &  0   & \\ \hline

\multirow{6}{5em}{Quadrotor} & Quadrotor Mass $m$ (kg) & 1  & .9 &
						 \\ & Inertias $J_x$ (kg m$^2$)& 8.1E-3 & 8.1E-3 &
						 \\ & $J_y$ (kg m$^2$) & 8.1E-3 & 8.1E-3 & $\int_{t_0}^{tf}(\mathbf{x}^TQ\mathbf{x} + \mathbf{u}^TR\mathbf{u})dt $
						 \\ & $J_z$ (kg m$^2$) & 14.2E-3 & 14.2E-3 &
						 \\ & Arm Length $L$ (m) & .24 & .24
\end{tabular}
\end{table}

\subsection{Cart Pole}
The first system is the classic cart pole. The equations of motion are given in Equation \eqref{CPEoM}, where the task is to swing up the cart pole from a downwards stable equilibrium to an unstable upwards equilibrium. Here $x$ is the position of the cart, $\theta$ is the angle of the pendulum with respect to the cart, and $f$ is the control force pushing the cart from side to side. The following equation is the true dynamics.

\begin{eqnarray}
\left[ \begin{array}{c} \ddot{x} \\ \ddot{\theta} \end{array} \right]  &=& \label{CPEoM}
\left[ \begin{array}{c}
\frac{-( b_2 \dot{\theta} \cos{\theta} - l f + l b_1 \dot{x} - m_p l \sin{\theta}(l\dot{\theta}^2 +  g\cos{\theta} )}{ l (m_c+m_p\sin^2{\theta}) } \\
\frac{-(-(b_2 \dot{\theta}(m_p + m_c) +  m_p l( b_1 \dot{x} - f)\cos{\theta} - m_p l\sin{\theta}((m_p + mc) g + m_pl\dot{\theta}^2 \cos{\theta})) }{l^2 m_p(m_c + m_p\sin^2{\theta})}
\end{array} \right]
\end{eqnarray}

We trained our Gaussian process where the input was the entire state vector and the control, and the output was $ \ddot{x}, \ddot{\theta}$. We generated 200 points of data using a sinusoidal desired trajectory from a crude feedback controller. However, even with this relatively small amount of data we were able to swing up, even though there was error in the desired position.

\begin{figure}[H]
	\centering
	\begin{subfigure}[b]{0.48\textwidth}
		\includegraphics[width=\textwidth]{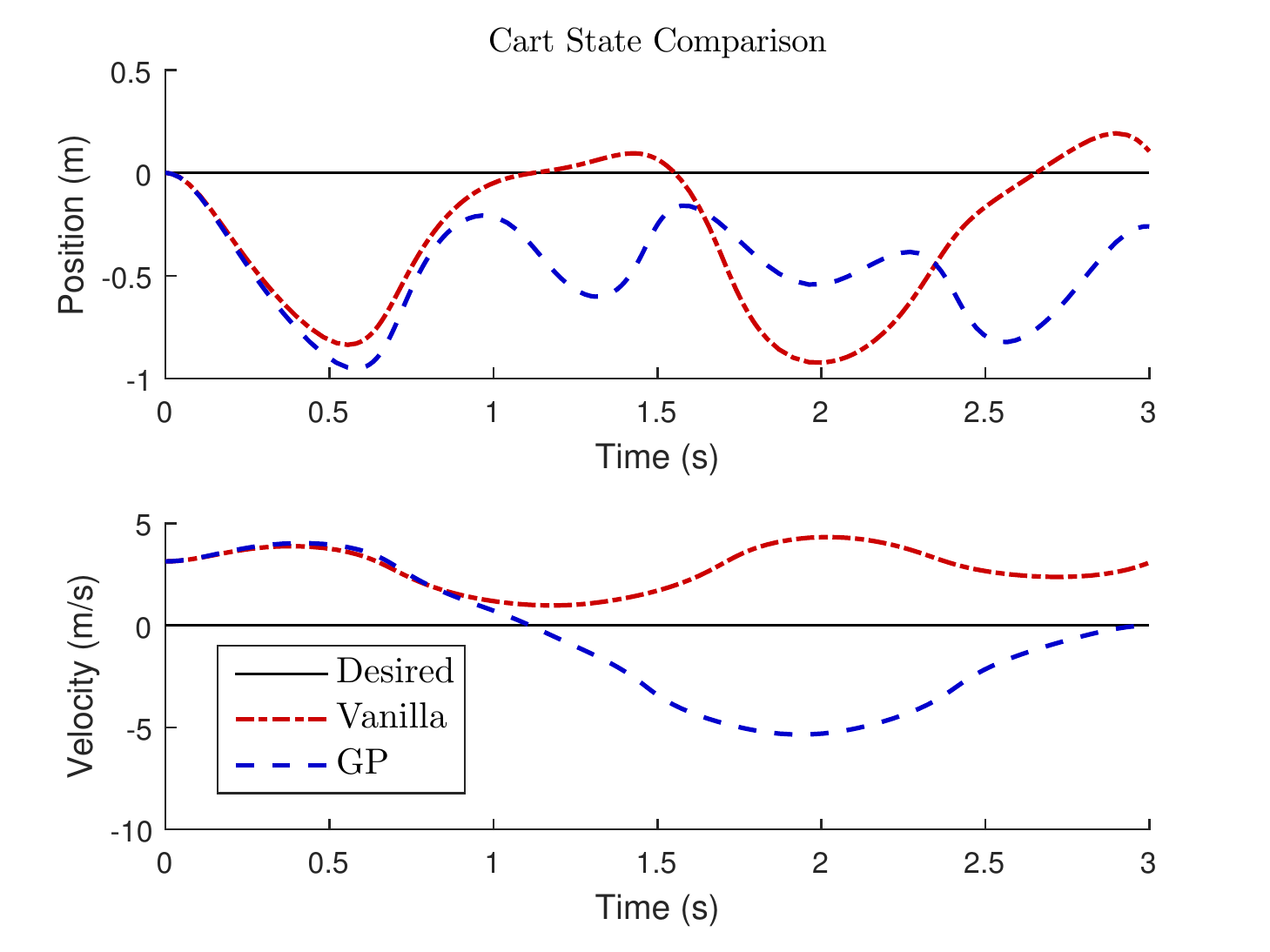}
		\caption{Position and Velocity}
	\end{subfigure}
	~
	\begin{subfigure}[b]{0.48\textwidth}
		\includegraphics[width=\textwidth]{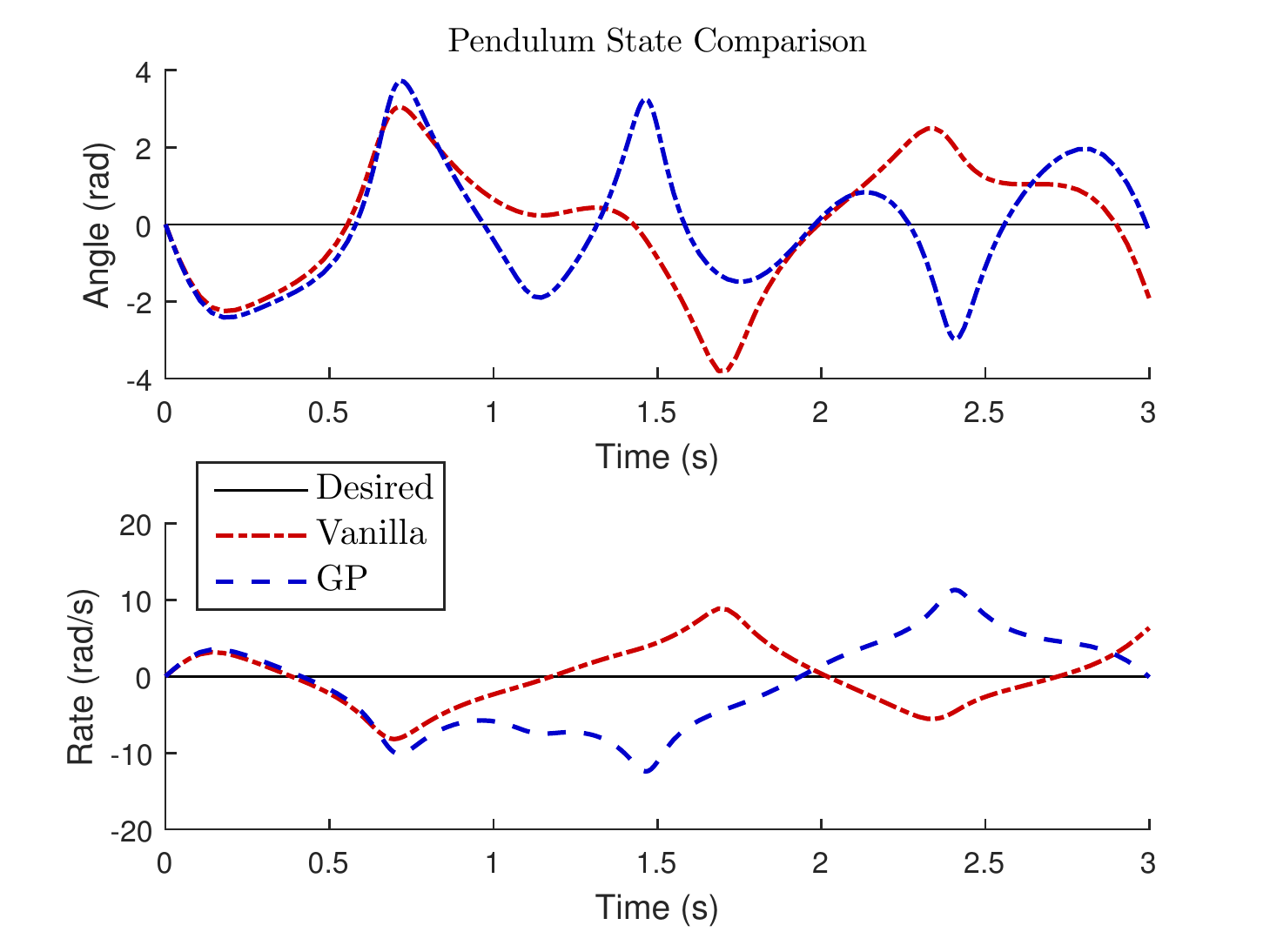}
		\caption{Angle and Angular Rate}
	\end{subfigure}
	\phantomcaption
\end{figure}
\begin{figure}[H]
\centering
\ContinuedFloat
	\begin{subfigure}[b]{0.48\textwidth}
		\includegraphics[width=\textwidth]{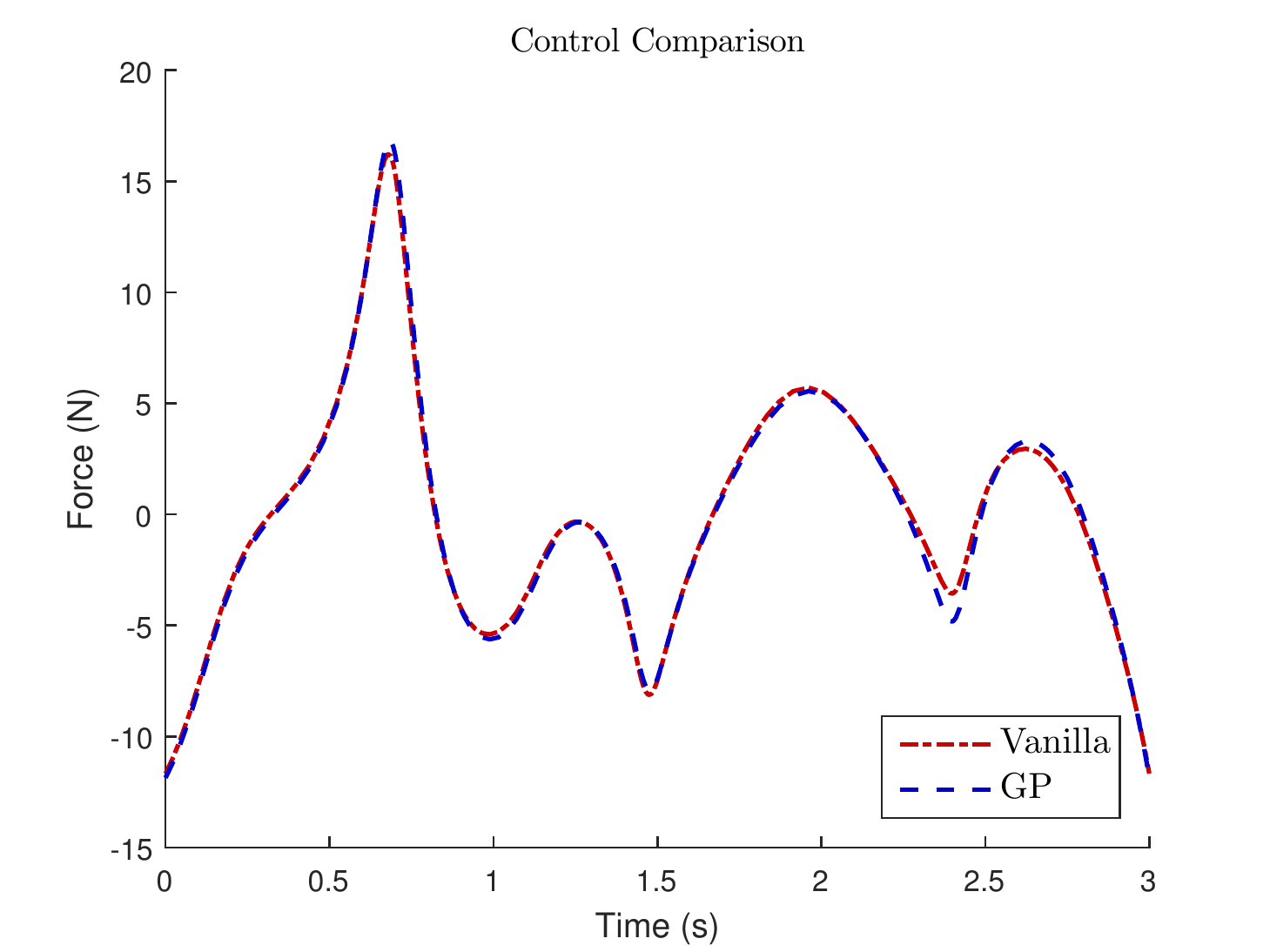}
		\caption{Control Effort}
	\end{subfigure}
	\caption{Episodic Cart Pole Comparison}
	\label{CP_van}
\end{figure}

We see that even minor unmodeled dynamics can have a drastic effect on the performance a nonlinear system in the episodic case. However, using this regression technique we were able to compensate. In the receding horizon case we see that performance is much improved, however the GP case is  marginally more accurate. Neither achieve exact convergence, illustrating a need for a more advanced algorithm.

\begin{figure}[H]
	\centering
	\begin{subfigure}[b]{0.48\textwidth}
		\includegraphics[width=\textwidth]{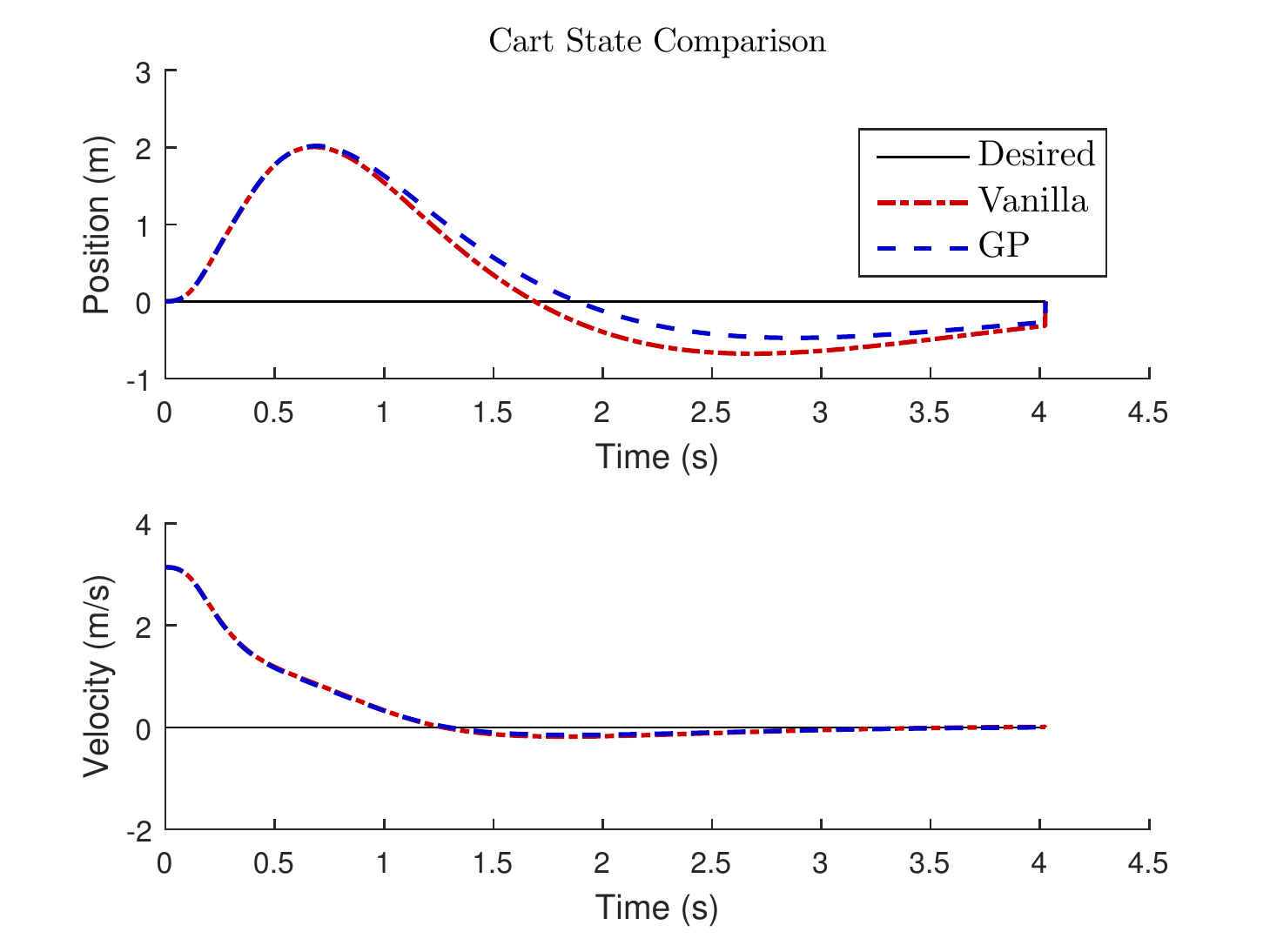}
		\caption{Position and Velocity}
	\end{subfigure}
	~
	\begin{subfigure}[b]{0.48\textwidth}
		\includegraphics[width=\textwidth]{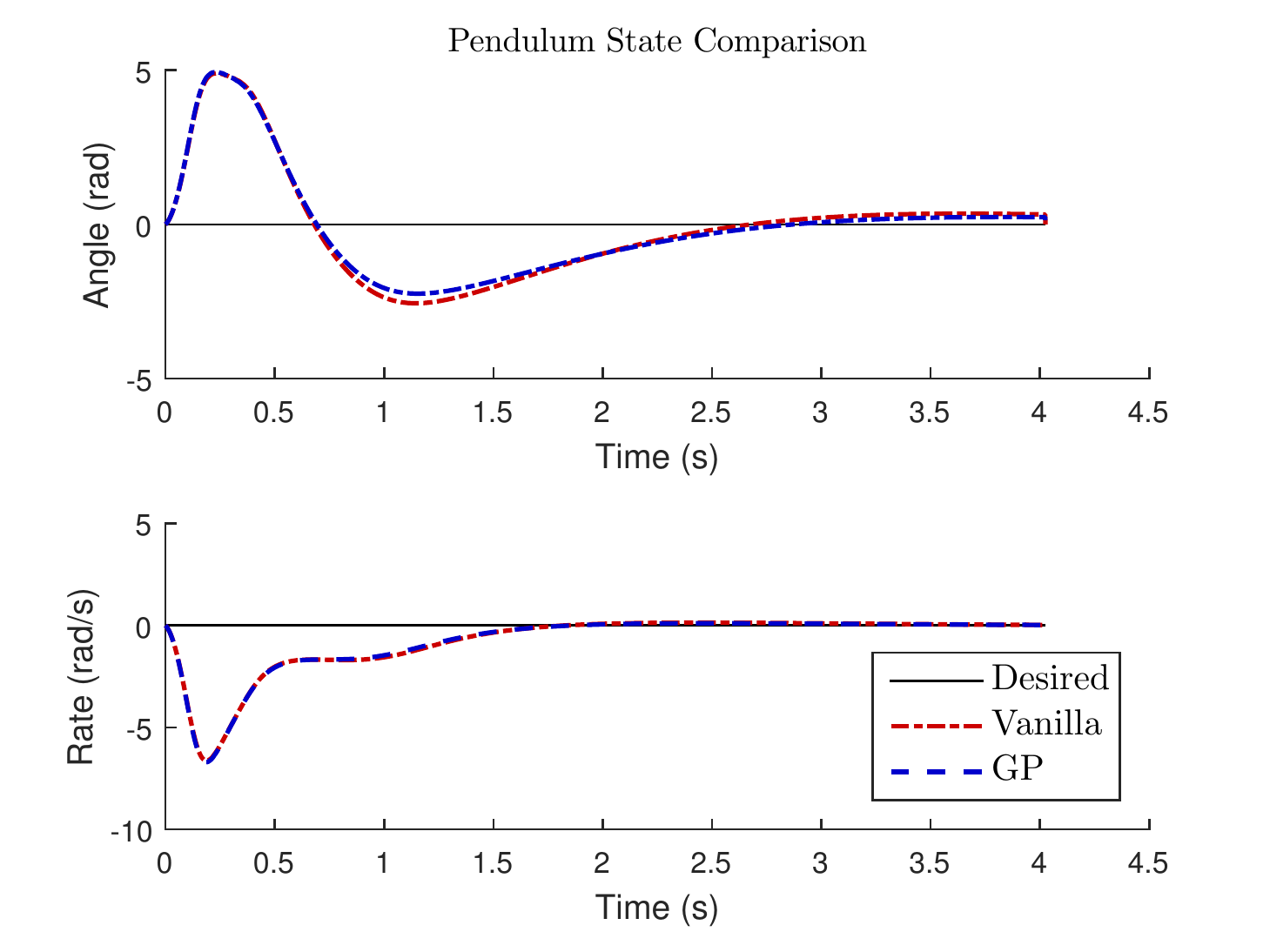}
		\caption{Angle and Angular Rate}
	\end{subfigure}
	\phantomcaption
\end{figure}
\begin{figure}[H]
\centering
\ContinuedFloat
	\begin{subfigure}[b]{0.48\textwidth}
		\includegraphics[width=\textwidth]{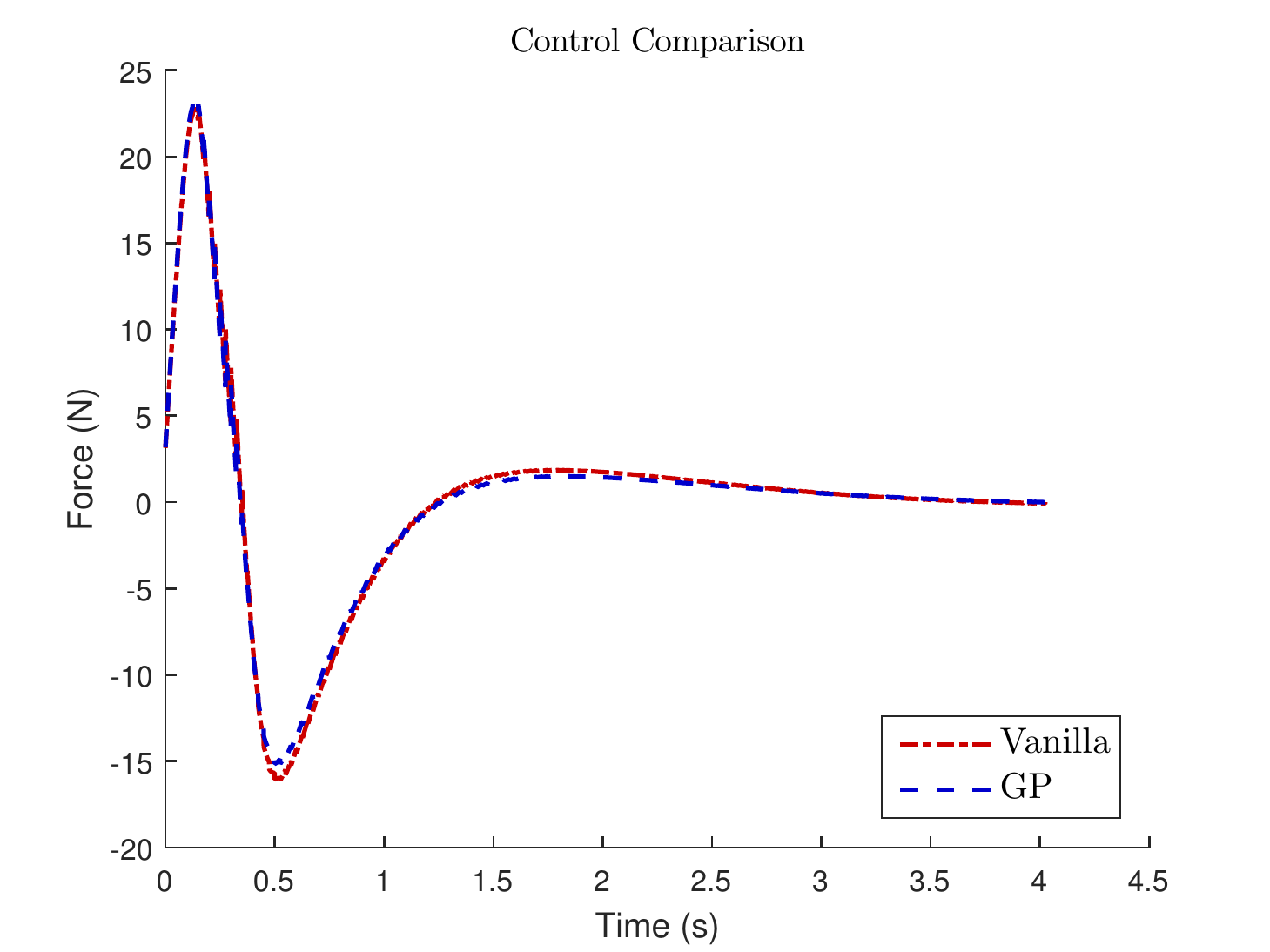}
		\caption{Control Effort}
	\end{subfigure}
	\caption{MPC Cart Pole Comparison}
	\label{CP_GP}
\end{figure}

\subsection{Quadrotor}

Now we explore the quadrotor, whose equations of motion are provided in Equation \eqref{QEoM}. The dynamical system has 12 states corresponding to position velocity, orientation, and angular rates. We use a simplified model of the quadrotor that assumes small deviations in the attitude. The position in 3D space is defined by $x$, $y$, and $z$, while the symbols $\phi$, $\theta $, and $\psi $ represent the roll, pitch, and yaw angles, respectively. There are now four controls in this system as opposed to singular control in the other systems, these four controls correspond to the thrust of each rotor. The initial position was $[-1,1,.5]$ and the target state was $[.5,-1,1.5]$. The cost function now has an additional term which penalizes excessive deflection of the pose, preventing the quadrotor system from rolling, pitching, or yawing large amounts to retain the validity of our model. In this case we did not introduce any new terms, we perturbed the mass of the quadrotor. The quad is defined WRT a north east down frame so a positive z indicates a downward movement.

\begin{eqnarray}
\left[ \begin{array}{c} \ddot{x} \\ \ddot{y} \\ \ddot{z} \\ \ddot{\phi} \\ \ddot{\theta} \\ \ddot{\psi} \end{array} \right] &=& \label{QEoM}
\left[ \begin{array}{c} 
\frac{1}{m}\big(\cos(\phi)\sin(\theta)\cos(\psi) + \sin(\phi)\sin(\psi)\big)\big( F_1 + F_2 + F_3 + F_4 \big) \\
\frac{1}{m}\big(\cos(\phi)\sin(\theta)\sin(\psi) - \sin(\phi)\cos(\psi)\big)\big( F_1 + F_2 + F_3 + F_4 \big) \\
g - \frac{1}{m}\big(\cos(\phi)\cos(\theta))\big( F_1 + F_2 + F_3 + F_4 \big) \\
\frac{1}{J_x}\big(\dot{\theta} \cdot \dot{\psi} (J_y - J_z) + L\big( F_4 - F_2 \big)\big) \\
\frac{1}{J_y}\big(\dot{\phi} \cdot \dot{\psi} (J_z - J_x) + L\big( F_1 - F_3 \big)\big) \\
\frac{1}{J_z}\big(\dot{\phi} \cdot \dot{\theta} (J_x - J_y) + 0.05L\big( F_2 + F_4 - F_1 - F_3 \big)\big)
\end{array} \right]
\end{eqnarray}

\begin{figure}[H]
	\centering
	\begin{subfigure}[b]{0.48\textwidth}
		\includegraphics[width=\textwidth]{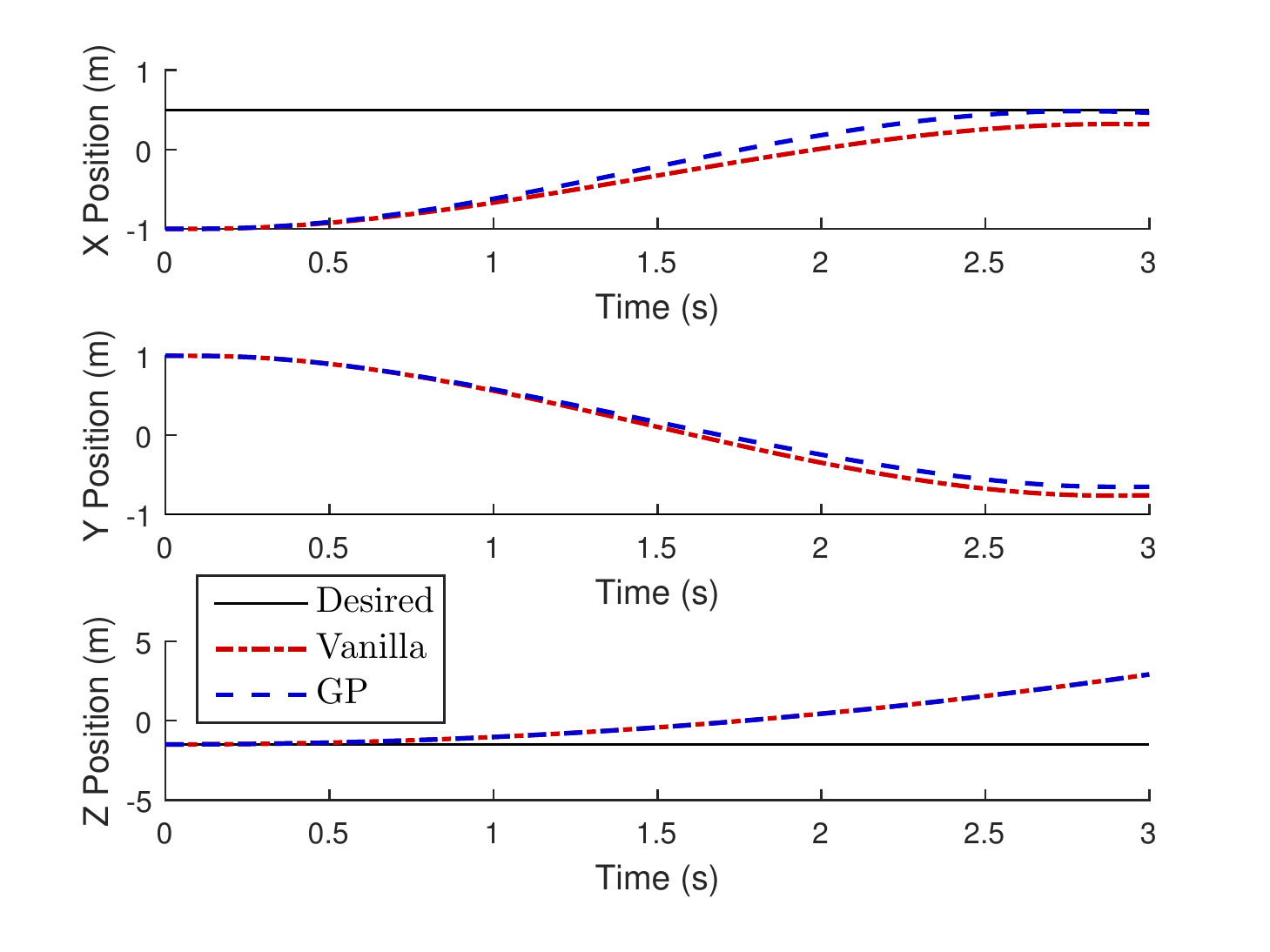}
		\caption{Position}
		\label{QP}
	\end{subfigure}
	~
	\begin{subfigure}[b]{0.48\textwidth}
		\includegraphics[width=\textwidth]{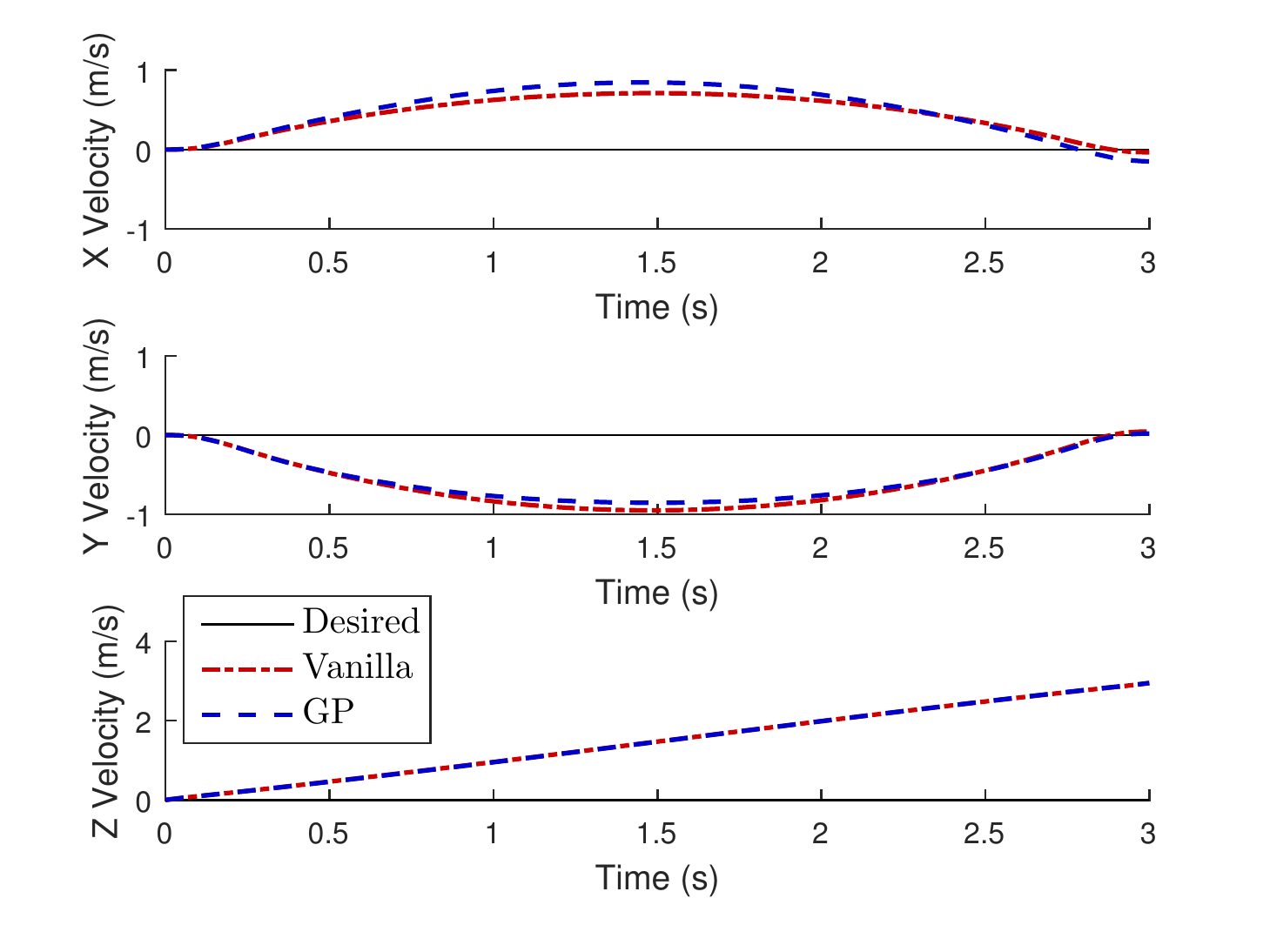}
		\caption{Velocity}
		\label{QV}
	\end{subfigure}
\\
	\begin{subfigure}[b]{0.48\textwidth}
		\includegraphics[width=\textwidth]{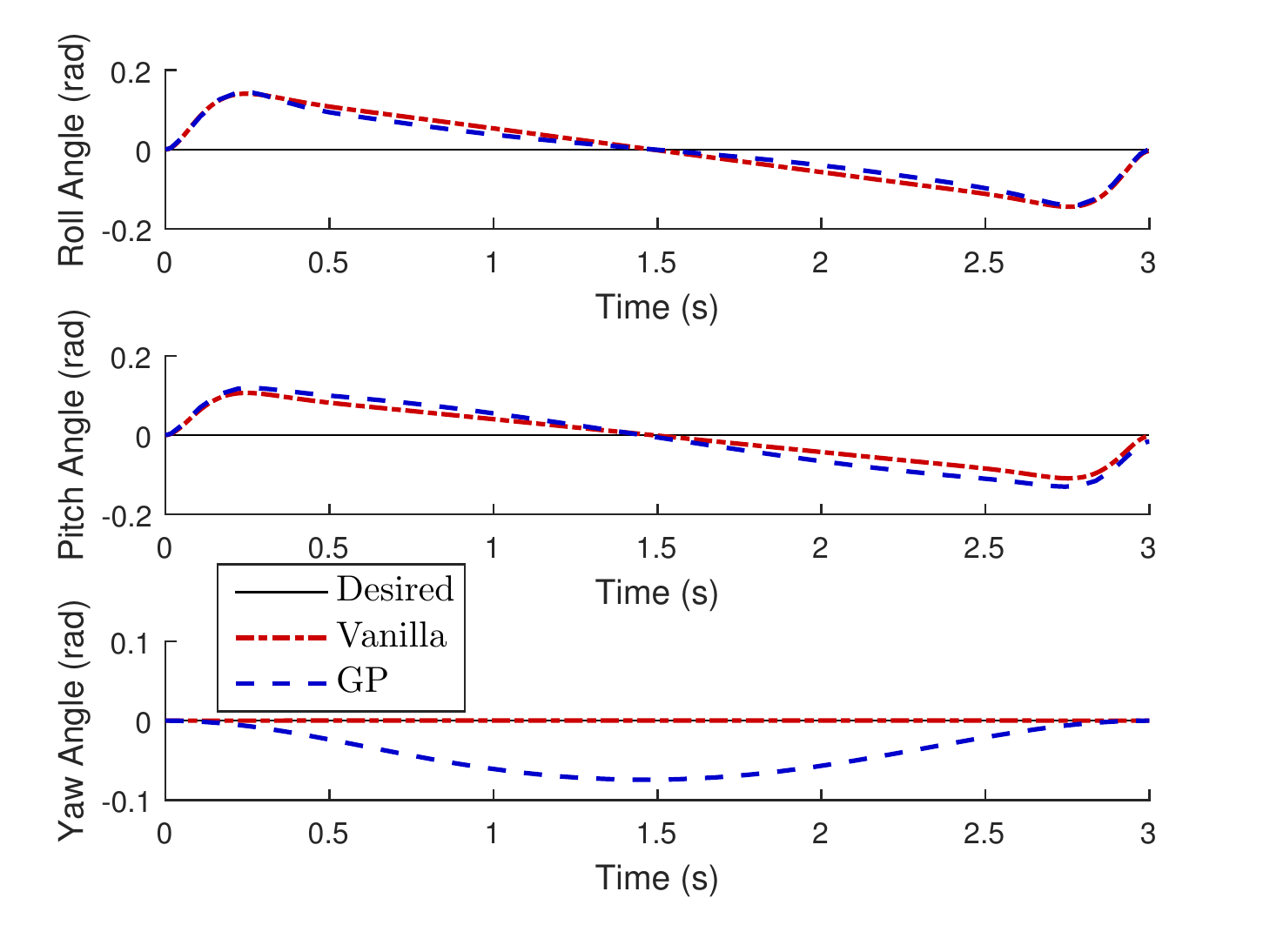}
		\caption{Pose}
		\label{QA}
	\end{subfigure}
	~
	\begin{subfigure}[b]{0.48\textwidth}
		\includegraphics[width=\textwidth]{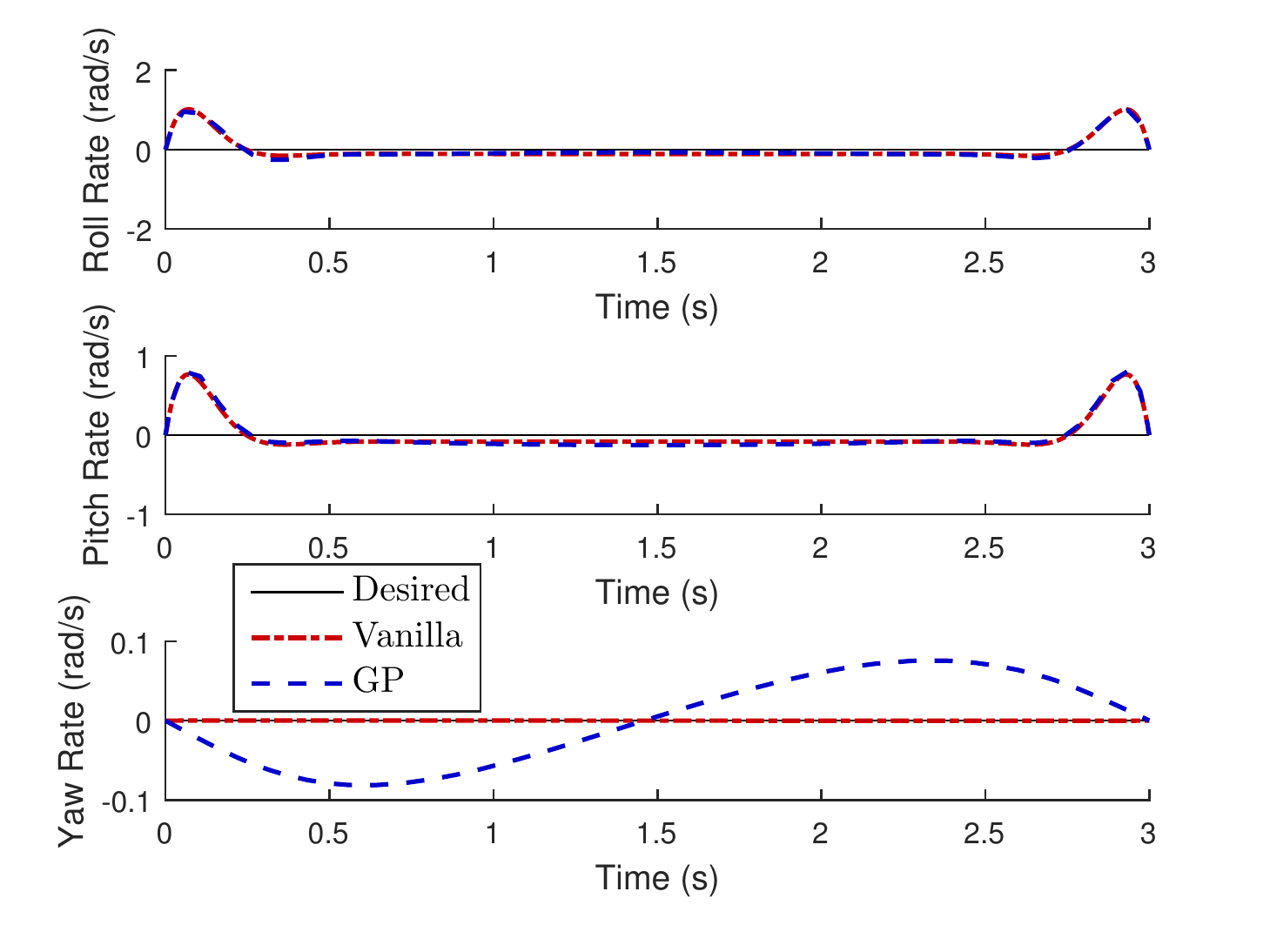}
		\caption{Angular Rate}
		\label{QR}
	\end{subfigure}
	\caption{Episodic Quadrotor Comparison}
\end{figure}

In this case, we see that there is a primary deviation in the state from the target in our z position. The Gaussian process  is trained from sinusoidal data in the x and y direction (i.e flying in circles) but had remained level in the z direction. We used 200 data points, but clearly the GP does not have enough data to make accurate predictions of the behavior in the z direction. The marginal difference between the nominal and semi-parametric dynamics further reaffirms the fact that our Gaussian Process is not contributing significantly due to a lack of data.

\begin{figure}[H]
	\centering
	\begin{subfigure}[b]{0.48\textwidth}
		\includegraphics[width=\textwidth]{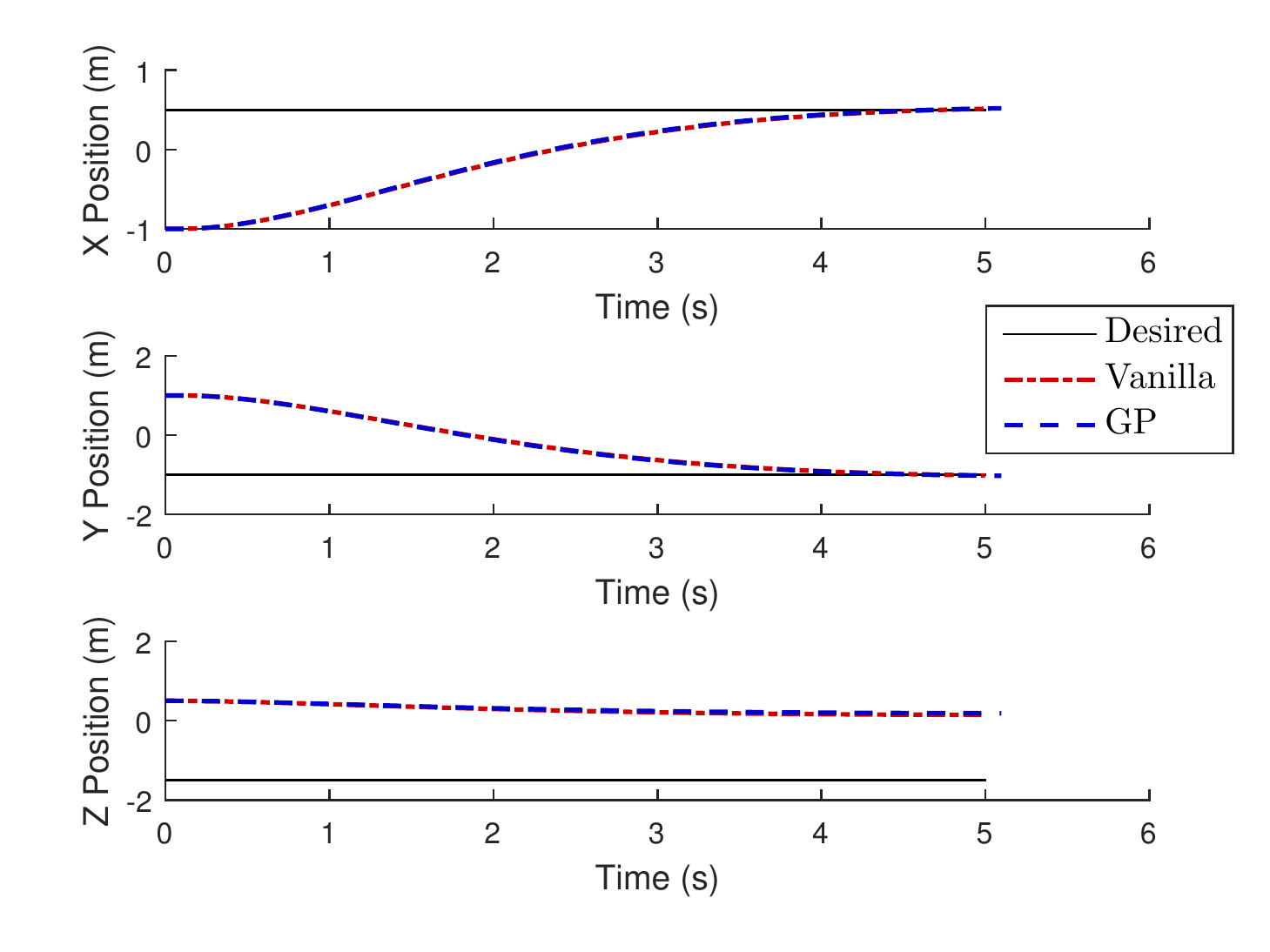}
		\caption{Position}
		\label{QP}
	\end{subfigure}
	~
	\begin{subfigure}[b]{0.48\textwidth}
		\includegraphics[width=\textwidth]{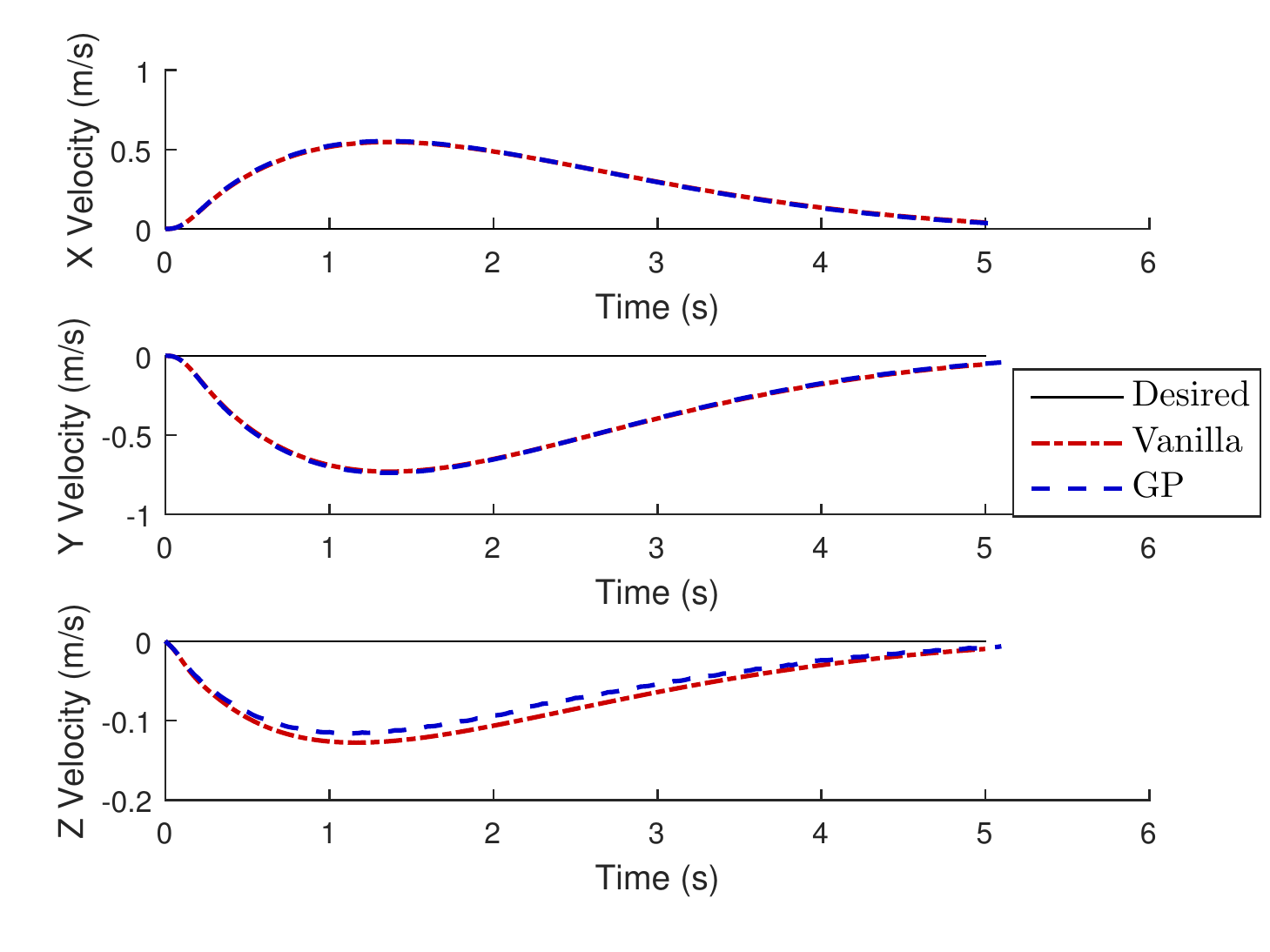}
		\caption{Velocity}
		\label{QV}
	\end{subfigure}
\\
	\begin{subfigure}[b]{0.48\textwidth}
		\includegraphics[width=\textwidth]{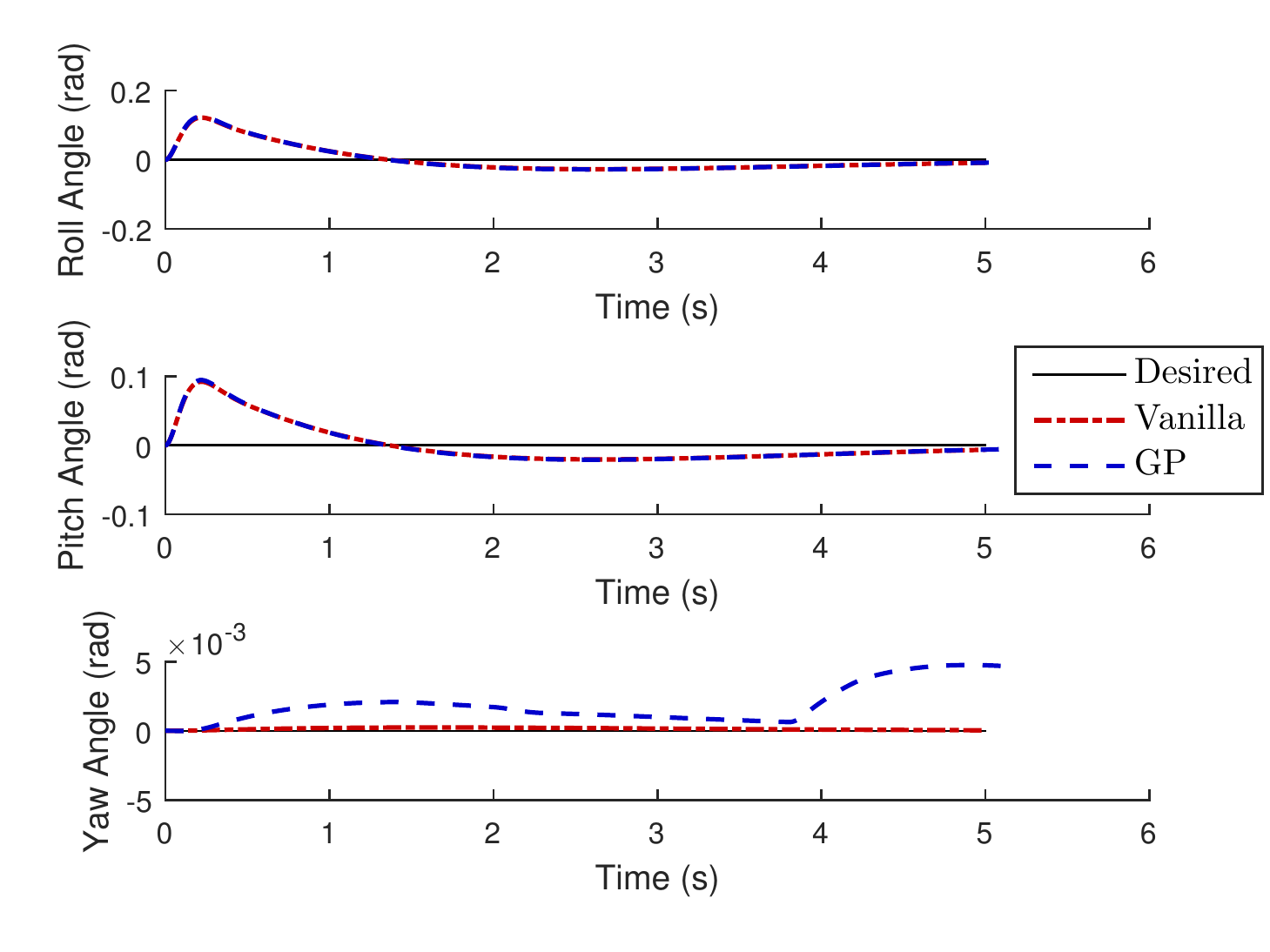}
		\caption{Pose}
		\label{QA}
	\end{subfigure}
	~
	\begin{subfigure}[b]{0.48\textwidth}
		\includegraphics[width=\textwidth]{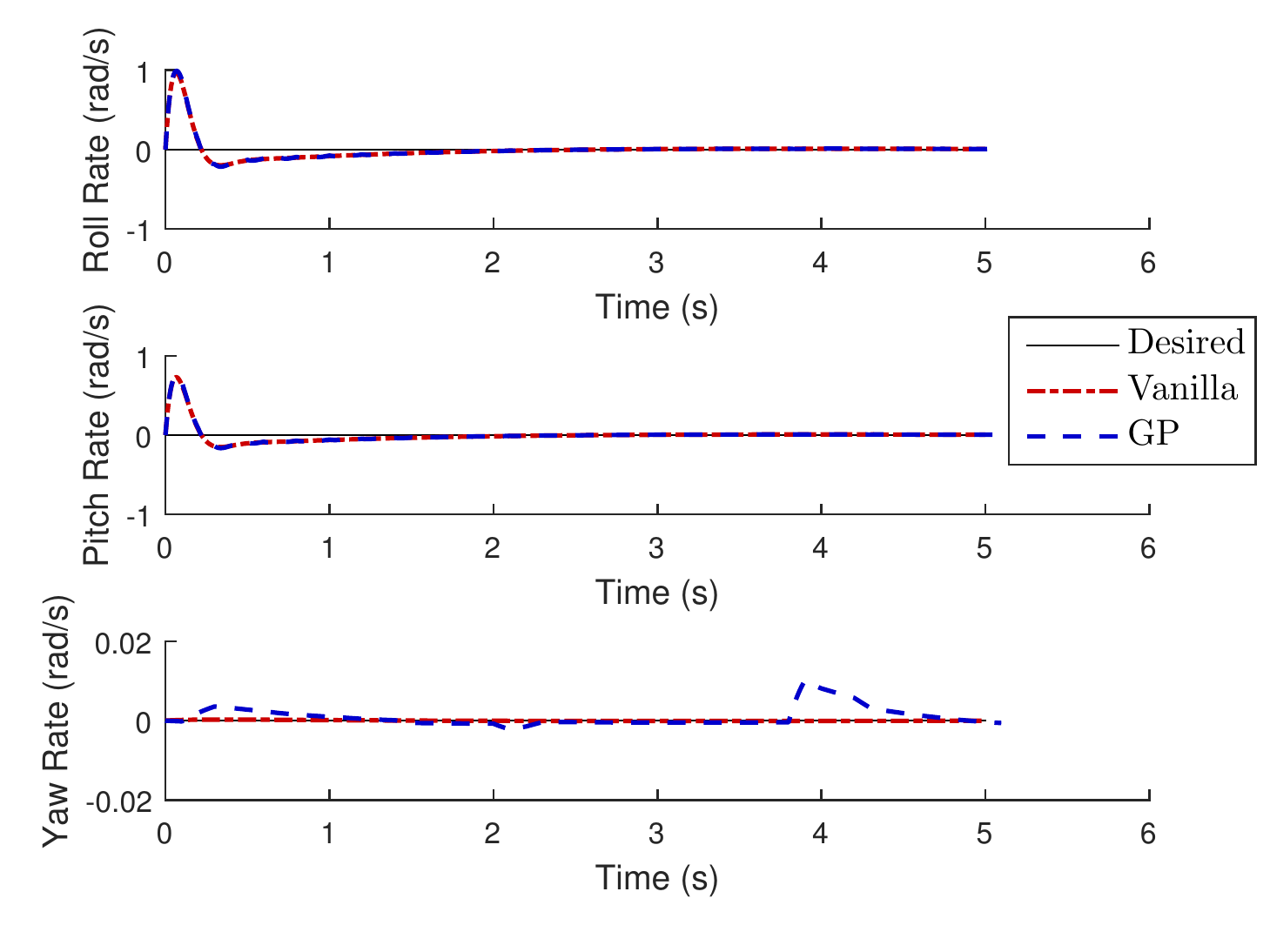}
		\caption{Angular Rate}
		\label{QR}
	\end{subfigure}
	\caption{MPC Quadrotor Comparison}
\end{figure}

In the MPC case, we again see that the z position of both systems are off the target, and the GP does not have enough data to compensate. As a result, another approach is required, where we can incorporate data as we perform rollouts of the trajectory optimizer. This is the reason we introduced Sparse Spectrum Gaussian Processes and Online Model Learning.

\subsection{Online Learning with Sparse Spectrum Gaussian Processes}

In this framework, we aim to update our Gaussian Process as we receive new data in order to improve performance. Note that since a Full Gaussian Process is data-driven, it retains all of the training data whenever a prediction is made. Thus adding more data online is not computationally efficient, as we must invert an $N$ dimensional kernel matrix each time we update. Sparse Spectrum GP's circumvent this problem by approximating this kernel matrix with randomly sampled Fourier features. Each data point that is an update to the SSGP constitutes a rank-1 matrix which makes the update computationally efficient. Note that we must optimize the hyperparameters of the system with an initial set of data. This means that we must have a representative set of the data that we will see, otherwise our predictions will be biased, even with the online learning. Next we present the algorithm for Model Predictive Pseudospectral Control with Online Learning:

\begin{algorithm}[H]
\caption{Model Predictive Pseudospectral Control with Online Learning}
\label{<your label for references later in your document>}
\begin{algorithmic}[1]
\STATE Initialize Sparse Spectrum Gaussian Process with initial training data
\STATE Initialize GPOPS-II with the necessary setup files
\WHILE{$|| x_0 - x_f|| < $tol and $t < t_{max}  $}
\STATE Run GPOPS-II with Initial State $x_0$, Target State $x_f$, and Time $t$
\STATE Rollout the first $p$ steps of the optimal control
\STATE Take the $p$ new data points and update the SSGP with equations (\ref{Aupdate}) and  (\ref{yupdate})
\STATE Reinitialize the guess for GPOPS-II with the previous optimization
\STATE Setup the next iteration of inputs for GPOPS-II
\ENDWHILE
\end{algorithmic}
\end{algorithm}

This simple addition to Model Predictive Pseudospectral Control will incorporate the new data as we roll it out. Currently, the issue with the implementation is that the SSGP must be defined and updated in a vectorized fashion in order for the computational to be practical. Since the number of calls to the SSGP is equal to the number of meshpoints, as the grid becomes more fine, we must perform inference many times. If the inference is done sequentially, then it is unfeasible to run this algorithm.

\subsubsection*{Cart Pole, and Quadrotor Simulations}

We modify the parameters further for the SSGP test to show the increased robustness of this algorithm. We increase the unmodeled damping for the cart pole system and add a one percent perturbation in the length of the quadrotor arm.

\begin{table}[H]
\centering
\caption{Online Learning Example Problem Settings}
\label{Problem}
\begin{tabular}{l | l | c | c| c}
Problem & Parameters & True & Nominal & Cost Function\\ \hline \hline
\multirow{6}{5em}{Cart Pole}& Cart Mass $m_c$ (kg)		&  1.0 & .9  &
						 \\ & Link Mass  $m_p$ (kg)		& 0.5  & 0.45  & $\int_{t_0}^{tf}(R\mathbf{u}^2)dt $
					 	 \\ & Link Length  $l$ (m)		& 0.5  & 0.45  & 
					 	 \\ & Ground damping $b_1$ (N/s) & 0.3  &  0   &
					 	 \\ & Pivot damping $b_2$ (Nm/s) & 0.3  &  0   & \\ \hline

\multirow{6}{5em}{Quadrotor} & Quadrotor Mass $m$ (kg) & 1  & .99 &
						 \\ & Inertias $J_x$ (kg m$^2$)& 8.1E-3 & 8.1E-3 &
						 \\ & $J_y$ (kg m$^2$) & 8.1E-3 & 8.1E-3 & $\int_{t_0}^{tf}(\mathbf{x}^TQ\mathbf{x} + \mathbf{u}^TR\mathbf{u})dt $
						 \\ & $J_z$ (kg m$^2$) & 14.2E-3 & 14.2E-3 &
						 \\ & Arm Length $L$ (m) & .24 & .2376
\end{tabular}
\end{table}

The following simulations were run with an in-house implementation of Pseudospectral Optimal Control (implemented in python) that implements the SSGP framework as a natural extension. The jacobian and hessian calculations were incompatible with ADIGATOR (the GPOPS-II autodifferentiation software). The following table has the nominal and perturbed parameters for both the cart pole and the quadrotor case. In this, the quadrotor has an additional arm length perturbation of one percent, and the cart pole has a much larger amount of unmodeled viscous friction. Again we see that without any compensation of for the dynamics difference, we the cart pole cannot swing up at all. However, using the SSGP online learning algorithm, the cart pole is able to swing up close to the equilibrium point, then actively stabilize itself.

\begin{figure}[H]
	\centering
	\includegraphics[width=0.75\textwidth]{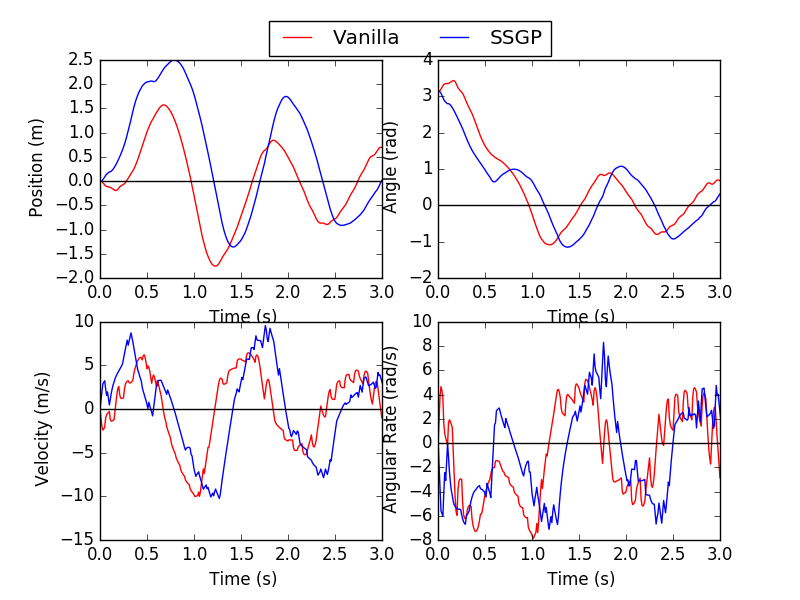}
	\caption{SSGP Receding Horizon Cart Pole Comparison}
\end{figure}

\begin{figure}[H]
	\centering
	\includegraphics[width=0.75\textwidth]{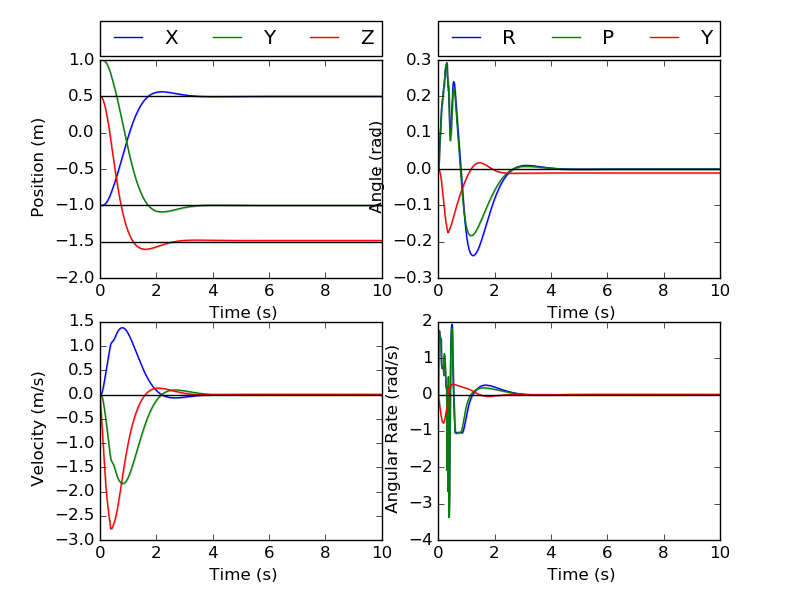}
	\caption{Quadrotor SSGP Comparison}
\end{figure}

For the quadrotor case the difference becomes even more drastic, as we see exact covergence with the SSGP algorithm fairly quickly. The vanilla case had an error in the height dimension which is clearly not present here.

\subsection*{Obstacle Avoidance using SSGP}

In our formulation of a semi-parametric representation, we use a probabilistic inference technique known as a Gaussian Process in order to represent unknown dynamics. A benefit of using this \textit{probabilistic} machine learning tool is that we can get an estimate of the uncertainty of our prediction. In this section we will show the equations to combine this uncertainty estimate along with the Model Predictive Pseudospectral Control (MPPC) to perform obstacle avoidance using hard state constraints. We can formulate the constraint with the following equation:

\begin{eqnarray}
C(x,u) = Distance - (O_r + B_r + U_r) \geq 0
\end{eqnarray}

This inequality constraint can be naturally incorporated into the pseudospectral framework, making obstacle avoidance an ideal application. The uncertainty radius is determined by the predictive covariance of the SSGP. We propagate the covariance forward using the following equation, then use this value as a one step prediction on the uncertainty of the system.

\begin{eqnarray}
\Sigma_{x_1} = \Sigma_{x_0} + \Sigma_{\dot{x_0}}dt^2 + 2 \Sigma_{x_0, \dot{x_0}}
\end{eqnarray} 

$\Sigma_{x_1}$ defines an ellipse around the obstacle body, using the angle between the body and the object, we can find the uncertainty radius $U_r$. We applied this on a dubin vehicle that uses bicycle dynamics, however for a deterministic system the performance of the with the variance was identical to performance without using the variance. We apply the technique on the following dynamics system excluding the $U_r$ term:

\begin{eqnarray}
\dot{x_1} = (1-b_1)u_1  \cos(x_3) \\
\dot{x_2} = (1-b_1)u_1  \sin(x_3) \\
\dot{x_3} = (1-b_1)u_2
\end{eqnarray}

The true values for $b_1$ and $b_2$ are 0.3, however the nominal vales are a Gaussian distribution $\mathcal{N}(0.3,0.3)$. Thus we can apply our algorithm over 50 trajectories and look at the distribution over states and over the path. We see that this constraint is adhered to at all points in the trajectory. The problem is to move from the point $(5,5)$ to the point $(2,-2)$, while avoiding an obstacle of radius $2$. 

\begin{figure}[H]
	\centering
	\includegraphics[width=0.75\textwidth]{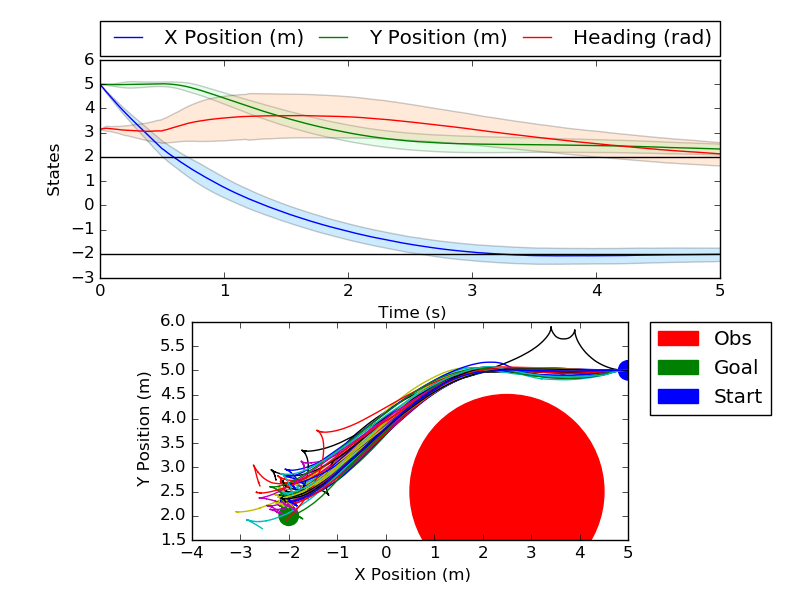}
	\caption{Dubin Vehicle Obstacle Avoidance}
\end{figure}

\section{Conclusion}
In conclusion, we see that even a small amount of viscous friction in the Cart Pole is enough to prevent an episodic trial of PSOC from reaching its target. The Gaussian process model is able to swing up the cart pole, however there is still error in the final position. In the MPC trials, both systems are able to swing up the cart pole and we see a marginal improvement using the Gaussian process. The reason why the GP model does not completely solve the issue is because it is data dependent, and if the Gaussian process input is far away from the training data, then the machine reverts back to the prior, in this case zero. These results are echoed in the quadrotor model. The Gaussian process is trained with sinusoidal tracking data from a MPC controller, however this data does not provide enough information about the z direction dynamics of the true system. It is unlikely that we can perfectly train a learned model a priori before we attempt a task. As a result, it is clear that we must go a step further and incorporate the new knowledge we gain from each rollout. A naive way of incorporating data would be to use the Full Gaussian Process at the expense of heavy computational load. Instead we opt for a Sparse Spectrum Gaussian process and present an algorithm to perform trajectory optimization with online learning. In these examples we see an improvement in the task completion for both the Cart Pole and Quadrotor systems, and we go a step further to perform obstacle avoidance in the Dubin Vehicle. The key points that must be stressed here is that performance of the algorithm hinges on the optimization of the Sparse Spectrum Gaussian Process. If there is not enough data, or if the hyperparameters of the data are not correct, then it is possible to have \textit{worse} performance than in the vanilla case. Overall this represents a powerful tool in the area of trajectory optimization, and as further work is completed, this algorithm can be implemented on real systems that can simultaneously perform path planning, system identification, and control.

\section*{Acknowledgments}
This work was was supported by the Department of Defense (DoD) through the National Defense Science \& Engineering Graduate Fellowship (NDSEG) Program.

\nocite{*}
\bibliographystyle{aiaa}
\bibliography{trbib}
\end{document}